\documentclass{article}

\usepackage{arxiv}

\usepackage[utf8]{inputenc} % allow utf-8 input
\usepackage[T1]{fontenc}    % use 8-bit T1 fonts
\usepackage{hyperref}       % hyperlinks
\usepackage{url}            % simple URL typesetting
\usepackage{csquotes}
\usepackage{booktabs}       % professional-quality tables
\usepackage{amsfonts}       % blackboard math symbols
\usepackage{nicefrac}       % compact symbols for 1/2, etc.
\usepackage{microtype}      % microtypography
\usepackage{lipsum}
\usepackage{fancyhdr}       % header
\usepackage{graphicx}       % graphics
\graphicspath{{media/}}     % organize your images and other figures under media/ folder
\usepackage{hyperref}
\title{NLP4Gov: A Comprehensive Library for Computational Policy Analysis}
\usepackage{geometry}
\usepackage{longtable}
\usepackage{tabularx, booktabs}
\usepackage{float}
\newcolumntype{s}{>{\hsize=.5\hsize}X}
\author{Mahasweta Chakraborti\\
  University of California, Davis \\
\texttt{mchakraborti@ucdavis.edu}
\And
Sailendra Akash Bonagiri\\
  University of California, Davis \\
\texttt{sbonagiri@ucdavis.edu}
\And
Santiago Virgüez-Ruiz\\
  University of Massachusetts Amherst\\
\texttt{svirguezruiz@umass.edu}
\And
Seth Frey\\
  University of California, Davis \\
\texttt{sethfrey@ucdavis.edu}}

\begin{document}  
\maketitle

\begin{abstract}

Formal rules and policies are fundamental in formally specifying a social system: its operation, boundaries, processes, and even ontology. Recent scholarship has highlighted the role of formal policy in collective knowledge creation, game communities, the production of digital public goods, and national social media governance. Researchers have shown interest in how online communities convene tenable self-governance mechanisms to regulate member activities and distribute rights and privileges by designating responsibilities, roles, and hierarchies. We present NLP4Gov\footnotetext[1]{\url{https://github.com/BSAkash/NLP4GOV}}, an interactive kit to train and aid scholars and practitioners alike in computational policy analysis. The library explores and integrates methods and capabilities from computational linguistics and NLP to generate semantic and symbolic representations of community policies from text records. Versatile, documented, and accessible, NLP4Gov provides granular and comparative views into institutional structures and interactions, along with other information extraction capabilities for downstream analysis.

\end{abstract}

\section{Introduction}

Online governance has lately drawn significant research interest in HCI and been extensively studied through their instantiations in collective knowledge generation \cite{hess2007understanding,Hess_2005,frischmann2014governing}, content moderation~\cite{matias2019civic,chandrasekharan_quarantined_2022,chandrasekharan_bag_2017}, crowd-sourcing and production of digital public infrastructure~\cite{gillespie2010politics}, including software critical to infrastructure at large~\cite{Schweik_English_2012,schweik2007tragedy,Schweik_Kitsing_2010,o2003guarding,eghbal2020working,Hippel_Krogh_2003}. Consequently, formal policy analysis is gaining traction in socio-technical systems research, particularly in how these collaborative initiatives regulate roles, rights, and responsibilities across the communities and beneficiaries involved.

Online communities are central in defining the virtual sphere and generating these goods and services of considerable economic value \cite{benkler2006wealth}. Moreover, they are also uniquely situated in their research significance, as circumstances motivating their emergent self-governance and mature policy systems are a microcosm of several other real-world challenges, particularly in public administration and natural resources management. Described as the 'Tragedy of the Commons" \cite{hardin1968tragedy}, natural resource conservation is especially complicated by 'free-riders' who harvest resources without reciprocal investments in production and upkeep. Policy functions to define resource and user boundaries, articulate sanctions, assign user rights and responsibilities, and overall implement sustainable management of finite resources~\cite{ostrom1990governing,Schlager1992PropertyRightsRA}. While digital goods are non-expendable, without adequate incentive structures and monitoring, collective interests, and benefits are similarly endangered by opportunistic users and limited contributions. Therefore, community policies are formal mechanisms formulated to mobilize participation and pool contributions, oversee maintenance and distribution of resources and goods generated, and also to monitor violations and other harmful behaviors \cite{frey2019emergence,schneider_modular_2021,frey2019place}. Lessons from self-sustaining communities may indeed carry implications for critical questions in public life and have intrigued researchers from fields beyond HCI, such as social science, anthropology, economics, and public policy~\cite{ostrom2009understanding,poteete2010working,forsman2021comparisons}. 

Significant contributions have been made lately in understanding the dynamics of governance activity, evolutionary trends, overlaps, and differences in policy content as well as their scope and themes, whether in online policy systems~\cite{fiesler2018reddit,10.1145/3584931.3606970} or in policy analysis generally~\cite{Siddiki_Heikkila_Weible_Pacheco‐Vega_Carter_Curley_Deslatte_Bennett_2022, siddiki2023understanding, siddiki2023evaluating, 4c419698-269e-3dde-8471-178d4492a08a, weible2011foes}. This has unlocked rich collections of governance documents, conversational corpus, and other archival footprints from communities. In order to derive key insights and patterns, empirical and data-driven researchers have explored computational methods to systematize and automate powerful and costly analytic approaches of policy analysis~\cite{rice_machine_2021, frantz2022institutional,ghorbani2014enhancing}. Yet relatively few scholars have pursued algorithmic approaches to semantically comprehend information from policy texts~\cite{chakraborti2023we, chandrasekharan_bag_2017}, or derive the underlying cognitive structures~\cite{artikis2009specifying,decaro2021motivational} and symbolic abstractions that bind institutional transactions~\cite{9459499,frey2023composing}. This is especially crucial as online collectivism continues to grow at an incredible pace, thus necessitating systematic, computationally scalable approaches to delve deeper into their governance behavior. NLP4Gov is a step to bridge this gap, as we extensively survey and string together concepts and approaches from computational linguistics and NLP to make computational policy analysis accessible to researchers across diverse backgrounds and questions. Built in a modular fashion while also supporting end-to-end data processing, NLP4Gov is intended to open up and demonstrate a range of possibilities for natural language understanding in the study of human behavior. It is built to assist data retrieval and provide detailed insights and visualizations into institutions, along with other information and measurements for downstream analysis.

\section{Related Work}
Considerable HCI and social media scholarship have explored regulatory patterns across online platforms~\cite{fish_birds_2011, gillespie2010politics}. Understanding the need for democratic governance and approaches for adaptive platform design has garnered focal interest with burgeoning user participation ~\cite{zhang_policykit_2020,schneider_modular_2021,tchernichovski_experimenting_2021,10.1145/3415178,10.1145/2441776.2441923,jhaver2023decentralizing}. Researchers have utilized advances in policy analysis, and increasingly available policy records and other governance instruments to characterize regulation approaches and their implications for community health and sustenance~\cite{Shaw_Hill_2014, 4c419698-269e-3dde-8471-178d4492a08a}. Multiple studies have discovered associations between governance complexity and growth among online communities \cite{frey_emergence_2019, frey_governing_2022,hill_hidden_2021}. Several studies centered their design around policy records or related corpus. Temporal analysis of major vernacular Wikipedias revealed similar patterns of governance activity yet variation in rule composition over time \cite{hwang_rules_2022}. Pater et al. found noticeable differences across multiple social media sites in their approach to harassment-related behavior \cite{pater_characterizations_2016}. In order to understand policy configurations across communities, Fiesler et al. codified rules written by Reddit moderators along their thematic and regulatory categories ~\cite{fiesler2018reddit}. Policy typologies \cite{Polski_Ostrom_1999} have been applied to constitutions from private trackers to study how pirate communities operate \cite{harris_institutional_2018}. Discourse from community threads has informed understanding of intellectual property protection in crowd-sourcing platforms \cite{bauer_intellectual_2016}. Chandrasekhar et al. examined posts and threads to understand the impact of platform moderation tactics across subreddits \cite{chandrasekharan_quarantined_2022}. In order to identify and mitigate abusive behavior online, content similarity-based approaches have been proposed that conserve human annotation and intuitively leverage existing big data traces for real-time governance \cite{chandrasekharan_bag_2017}. Recent work in open source management has studied interrelationships between rules invoked by OSS developers in their daily operations and the sociotechnical evolution of their communities \cite{yin2022open}. More recently, Chakraborti et al. studied formalization among OSS developers through email records by measuring routine activities described and their semantic internalization of written, established policies~\cite{chakraborti2023we}. These scholastic developments open promising directions for big-data corpus in governance analysis and consolidates the motivation for a comprehensive kit to support cross-platform research and interdisciplinary convergence supported by state of art natural language processing. 
 \section{Setup and Usage}

The NLP4Gov repository comprises a collection of interactive applications developed on Jupyter notebooks hosted through Google Colaboratory. We opted for Colaboratory for its dedicated, interactive environments, usability, and legibility for both proficient programmers and researchers with limited computational training. Importantly, it supports a range of self-contained platform-agnostic demonstrations, permits runtime requests to personal or external institutional systems, poses minimal setup requirements, and allows upgrades to storage and compute capabilities (through Google's own infrastructure) specific to research needs. The basic free version itself comes with CPU/GPU sufficient to execute the models (also the library defaults) and produce the results we describe in the current work. 

The back-ends use transformer-based language models that were fine-tuned on standardized NLP benchmarks and further adapted by us for end-to-end policy analysis. The repository is designed to be amenable to researchers across disciplines and does not require extensive development experience for use. The notebooks simply require users to execute a small set of inline commands to load requisite scripts and obtain inferences on their data. Additionally, we provide extensive documentation on how to format data for specific applications and navigate the Colaboratory interface. We include ample datasets from different online communities such as Reddit ~\cite{frey_governing_2022}, and OSS foundations like Apache \cite{Sen_Atkisson_Schweik_2022, yin2021apache}, and also data from public policy case studies \cite{siddiki2023understanding,siddiki2012using,siddiki2014assessing} to extend general application, experimentation, and evaluation for researchers to assess how the applications may serve their goals. 

 \section{Applications and Pipelines}
  Researchers have analyzed policy content to understand the structural features of governance systems or discern similarities and uniqueness of regulatory patterns. Table ~\ref{tab:apps}. provides an overview of the six major applications we have currently developed. Each of these can be used independently or cascaded into pipelines. Subsequent sections further detail their development and potential analytical approaches they can support. Researchers interested in adapting the applications to more specific constructs of interest or wishing to incorporate domain-specific/newer language models may additionally modify the source to their requirements.   
  
 \begin{table}[!ht]
 \centering
 \begin{tabularx}{\columnwidth}{
  |p{\dimexpr.4\columnwidth-2\tabcolsep-1.3333\arrayrulewidth}% column 1
  |p{\dimexpr.6\columnwidth-2\tabcolsep-1.3333\arrayrulewidth}|}
        \hline
        \hfil Application & \hfil Description \\ \hline
        \hfil ABDICO\_coreferences & Substitutes determiners and other coreferences \cite{jurafsky2000speech,lee-etal-2018-higher} with the actual named entity over document sections \\ 
        \hline
        \hfil ABDICO\_parsing & Institutional Grammar \cite{crawford_grammar_1995} parsing from individual policy statements
        \\ \hline
        \hfil ABDICO\_clustering & Clustering / topic modeling of policies or their components using Bertopic \cite{grootendorst2022bertopic} and semantic embeddings \cite{Reimers_Gurevych_2019}
        \\ \hline
        \hfil Policy\_comparison & Semantic comparison of policies across community databases 
        \\ \hline
        \hfil Policy\_explore & Data retrieval with policies / governance topics from trace data / community records
        \\ \hline
        \hfil Reddit\_governance & Interactive policy comparison across popular subreddits \cite{frey_governing_2022}
        \\ \hline
        % Posts that omit essential information, or present unrelated facts in a way that suggest a connection will be removed. & r/todayi learnt & Use only original, reliable sources for your articles. Sites that frequently rehost stories are on the [Autoremoval List](https://www.reddit.com/r/nottheonion /wiki/autoremoval). Blogs, tabloids, activist pages or satire websites are not reliable sources & r/notthe onion & 0.43 \\ 
        % \hline
        % Community & Rule & Community & Rule & Similarity Score \\ 
        % \hline
        %  \hline
    \end{tabularx}
    \caption{Summary of current applications and usage}
 \label{tab:apps}
 \vspace*{-10pt}
\end{table}

 \subsection*{Comparative Policy Analysis}
 \textit{Policy\_comparison} explores semantic approaches to assess the similarity of policy texts at the conceptual level~\cite{jurafsky2000speech}, and facilitates comparative analysis of governance. Classical methods such as regular expressions or lexical matching, while extremely powerful, may sometimes be limited in rating equivalence (or lack thereof) between complex concepts articulated by policies. This is particularly true when related ideas are expressed through synonyms or statements with high word overlap yet bearing different implications.

The \textit{Policy\_Comparison} application builds on language models that interpret and encode text based on the context jointly conveyed by words within the sentences. They were trained in a manner such that conceptually similar statements are also encoded into representations closer to each other. For any given pair of policy databases, we generate a score for the mutual similarity between all pairs of policies through the cosine similarity between their encoded text embeddings. For encoding policies, we use a bi-encoder model \cite{Reimers_Gurevych_2019} built on the MPNET architecture~\cite{song2020mpnet}, which was trained on multiple datasets for broad, multi-domain usage\footnote{https://www.sbert.net/docs/pretrained\_models.html, accessed 03/20/2024}. The final result presents a list of policy pairs ranked in descending order of similarity.  

 Table ~\ref{tab:compol}. shows examples of rules and their mutual similarity produced using \textit{Reddit\_governance}. This demo app demonstrates Policy\_comparison over the top 100 most popular of all the subreddits studied by Frey et al. \cite{frey_governing_2022}. Users may choose any pair of communities from a webform in the notebook to compare their policy databases. Having computed these similarities, a researcher could potentially use them to link related institutions or study community attributes explaining such similarities/dissimilarities.

 \subsection*{Tracking Governance in Action}

 Practical incidents reported and deliberated by members are invaluable for tracking the impact and evolution of governance systems. While policies are generally concise statements, narratives from community threads and discussions are often longer, more expressive, and uniquely informative of lived realities. \textit{Policy\_explore} enables one to host their own search engine and discover interactions relevant to a policy or concept of interest. Also based on a semantic approach, this application pursues principles similar to \textit{Policy\_comparison}. However, it employs asymmetric search and performs a slightly elevated task of efficiently comparing texts of dissimilar lengths, i.e., policies and conversations. Therefore, the application incorporates models designed to encapsulate wider contexts for comparison with the search query. In Policy\_explore, we use a Sentenceformer~\cite{Reimers_Gurevych_2019} model built on Distilbert \cite{sanh2019distilbert} and trained on MSMARCO \cite{nguyen2016ms} for document and passage retrieval capabilities. The search query can either be a phrase (e.g., \textit{"Content Moderation"}) or a full policy statement, and the application retrieves all posts/exchanges ranked in decreasing order of semantic relevance to the query. Example demonstrations walk users through retrieval of full-text emails related to policies from OSS community lists~\cite{chakraborti2023we, yin2021apache} in the Apache Software Foundation.   
 
 \textit{Policy\_explore} is expected to supplement multiple research directions. Semantic data retrieval can help researchers focus on relevant data subsets related to governance, thus streamlining coding efforts. Case studies of community behavior across their relatedness to policy may provide rich insight into pertinent questions, such as their changing scope or interpretation, conditions that perpetrate exceptions/violations, and how communities address such challenges. 

\begin{center}
 \begin{table*}[!ht]
 
 \begin{tabularx}{.99\linewidth}{
  |p{\dimexpr.30\linewidth-2\tabcolsep-1.3333\arrayrulewidth}% column 1
  |p{\dimexpr.12\linewidth-2\tabcolsep-1.3333\arrayrulewidth}% column 2
  |p{\dimexpr.30\linewidth-2\tabcolsep-1.3333\arrayrulewidth}% column 2
  |p{\dimexpr.12\linewidth-2\tabcolsep-1.3333\arrayrulewidth}% column 2
  |p{\dimexpr.15\linewidth-2\tabcolsep-1.3333\arrayrulewidth}|% column 2
  }
        \hline
        \hfil Rule & \hfil Community & \hfil Rule & \hfil Community & \hfil Similarity Score \\ \hline
        Medical advice is strictly prohibited on AskScience. Asking for or soliciting medical advice are both against the rules. & r/Ask Science & Offering or seeking medical advice is strictly prohibited and offending comments will be removed. Discussions regarding the advantages and/or disadvantages of certain treatments, diets, or supplements are allowed as long as relevant and reputable evidence is provided. & r/Science & \hfil 0.67 \\ 
        \hline
        AskScience has a strict policy against abusive and offensive language. Unless that language is in the context of research, it has no place here. We hold comments and posts to a high level of professionalism. We require our users and volunteers to always maintain a level of professionalism in order to participate. & r/Ask Science &  * Threats, suggestions of harm, personal insults and personal attacks are prohibited & r/sports & \hfil 0.53 \\ 
        \hline
        % Posts that omit essential information, or present unrelated facts in a way that suggest a connection will be removed. & r/todayi learnt & Use only original, reliable sources for your articles. Sites that frequently rehost stories are on the [Autoremoval List](https://www.reddit.com/r/nottheonion /wiki/autoremoval). Blogs, tabloids, activist pages or satire websites are not reliable sources & r/notthe onion & 0.43 \\ 
        % \hline
        % Community & Rule & Community & Rule & Similarity Score \\ 
        % \hline
        %  \hline
    \end{tabularx}
    \caption{Examples of most related policies detected by relative cosine similarity between subreddits with similar and dissimilar interests}
 \label{tab:compol}
 \vspace*{-10pt}
\end{table*}
\end{center}

 \subsection*{Institutional Analysis}
 \subsubsection{Institutional Grammar Framework}

Researchers have explored systematic approaches to represent the interdependencies spanning organizations and communities through the granular decomposition of the policies binding them. Notable among them is the Institutional Grammar (IG) \cite{crawford_grammar_1995}, a comprehensive framework that takes a syntactic approach to decompose policy texts into granular units. It defines an institutional statement or a self-contained policy sentence as the most fundamental unit of policy analysis. The updated IG 2.0 specification \cite{Siddiki_Heikkila_Weible_Pacheco‐Vega_Carter_Curley_Deslatte_Bennett_2022,frantz_institutional_2021} lays down the taxonomy of policy constituents, with four core components being the attribute, object, aim, deontic, context, and or else. The aim (I) or the main verb of the statement specifies the central actions or goals of the institutional statement. The attribute (A) or agent is usually the grammatical subject and represents the individuals or organizations who are required to execute the aims outlined by a policy. The object (B) of an institutional statement (generally the grammatical object) spans other entities/organizational devices that are the target recipients of the policy aim. The deontic (D) conveys the strength of an institutional statement. It's the prescriptive component that states the extent to which the particular regulation is binding, typically through modals such as may, can, should, must, and shall, etc. Our library currently supports the extraction of these four components from text data, as these are most fundamental to discovering existing interaction networks and power structures within an institution. 

\subsubsection{Parsing Policy Constituents}
\label{ig_parsing}
  There has been significant attention in recent years on automating IG parsing to draw insights from extensive collections of policy data. Vannoni \cite{vannoni2022political} performed IG analysis through dependency parsing. Rice et al. \cite{rice_machine_2021} identified ABDICO components using neural network classifiers over dependency features. We hereby introduce our approach, which meticulously combines dependency structures and semantic role labeling (SRL). While dependency parsing identifies grammatical tags of words, semantic role labeling/SRL is a computational linguistics approach that selects words/text spans within sentences that describe actor-objects, as well as conditions associated with a predicate (verb) in a given sentence. SRL comprises a set of core arguments, numbered ARG0-ARG5 and ranked in order of agentative precedence. ARG0 is generally the main agentative argument except for intransitive verbs called unaccusatives~\cite{bonial2010propbank}, in which case the verb's agent is annotated as ARG1. The remaining core arguments (after the agent) represent direct/indirect objects. The modals describing the verb and negation (e.g., "shall \textit{not} submit a proposal") are classified as ARGM-MOD and ARGM-NEG, respectively. 
  
  We use pre-trained SRL models developed and validated on standardized datasets \cite{weischedel2013ontonotes} to facilitate off-the-shelf use for IG parsing. Shi et al.\cite{shi2019simple} developed a BERT-based approach to parse semantic roles for every verb in a sentence. Since the aim concerns the central activity being regulated by an institutional statement, we assign the ROOT verb of a statement's dependency tree to the Aim using Stanza \cite{qi2020stanza}. In cases where the ROOT of the statement is not a verb (e.g., certain subordinating clauses), we treat the verb with the most extensive SRL parsing (spanning most semantic roles and dependencies) as the aim. Next, we treat the Aim as the anchor and map other IG constituents to their semantic roles. Table ~\ref{tab:srlmap}. summarizes our approach. In cases where ARG0 is not present but the aim otherwise points to a direct subject, we treat ARG1 as the agent.

\begin{table}[ht]
\centering
 \begin{tabularx}{\columnwidth}{
  |p{\dimexpr.3\columnwidth-2\tabcolsep-1.3333\arrayrulewidth}% column 1
  |p{\dimexpr.7\columnwidth-2\tabcolsep-1.3333\arrayrulewidth}|}
 % {
 %  |p{\dimexpr.2\linewidth-2\tabcolsep-1.3333\arrayrulewidth}% column 1
 %  |p{\dimexpr.8\linewidth-2\tabcolsep-1.3333\arrayrulewidth}|% column 2
 %  }
        \hline
   ABDICO  Constituent & \hfil SRL-IG mapping \\\hline
   Aim (I) & ROOT of dependency tree (if verb) or predicate of the longest SRL graph \\\hline
   Attribute (A) & ARG0 of Aim if present else ARG1 if the statement has a nominal subject. None otherwise \\\hline
   Object (O) & Second highest core argument after the Attribute, in order of ARG1-ARG5 \\\hline
   Deontic (D) & ARGM-MOD + ARGM-NEG \\\hline

    \end{tabularx}
    \caption{Rule-based framework for combining Semantic Roles and Dependency Parsing for Institutional Grammar}
\label{tab:srlmap}
\vspace*{-10pt}
\end{table}

 \begin{table}[!ht]
 \centering
 \begin{tabularx}{\columnwidth}{
  |p{\dimexpr.38\columnwidth-2\tabcolsep-1.3333\arrayrulewidth}% column 1
  |p{\dimexpr.17\columnwidth-2\tabcolsep-1.3333\arrayrulewidth}% column 2
  |p{\dimexpr.15\columnwidth-2\tabcolsep-1.3333\arrayrulewidth}% column 2
  |p{\dimexpr.15\columnwidth-2\tabcolsep-1.3333\arrayrulewidth}% column 2
  |p{\dimexpr.15\columnwidth-2\tabcolsep-1.3333\arrayrulewidth}|% column 2
  }
        \hline
        Dataset & Attribute (F1) & Object (F1) & Aim (F1) & Deontic (F1) \\ \hline
        Food Policy Data \newline (N = 398) & 0.71 &  0.57 &  0.82 &  0.94   \\ 
        \hline
        % Food Policy Data (Rice et al.) & 0.62 & \textbf{0.75} &   \textbf{0.85} & \textbf{0.97} \\ 
        % \hline
        Aquaculture Policy \newline (N = 153) & 0.76 &  0.57 &  0.81 & 0.93  \\ 
        \hline
         National Organic  \newline Policy (N = 835) & 0.86 & 0.49 & 0.84 & 0.94  \\ 
        \hline
        % Community & Rule & Community & Rule & Similarity Score \\ 
        % \hline
        %  \hline
    \end{tabularx}
    \caption{Performance of Semantic Role Labeling for parsing Institutional Grammar Constituents}
 \label{tab:val}
 \vspace*{-10pt}
\end{table}

Table ~\ref{tab:val}. reports the performance of our framework in terms of the word-constituent match on multiple datasets that were manually coded for IG constituents. Statements may often have the same modals governing multiple verbs, as a result the detection of Deontics is often unaffected by the exact predicate we chose as Aim or anchor for our SRL-IG mapping. Rice et al. reported performances higher for objects (0.75 F1), lower on attributes (0.62 F1), and comparable for aims and deontics on the Food Policy (FPC) dataset \cite{rice_machine_2021}, a frequent benchmark used in IG-based studies. Despite some differences in our respective validation approaches\footnote{While we report performance on the full dataset without further training, Rice et al. developed and validated their approach specifically on the FPC dataset.} This reference highlights the need to improve our performance for extracting Objects.  In order to assess methodological generalizability across different data distributions, we performed additional validation with two other datasets. We find similar trends in performances across components for all three cases, with room for further improvement in Objects. Further details on the datasets and processing are available in Appendix ~\ref{validation}. 

\begin{figure}[htbp] 
\centering
\includegraphics[width=0.99\textwidth,height=.7\textheight,keepaspectratio]{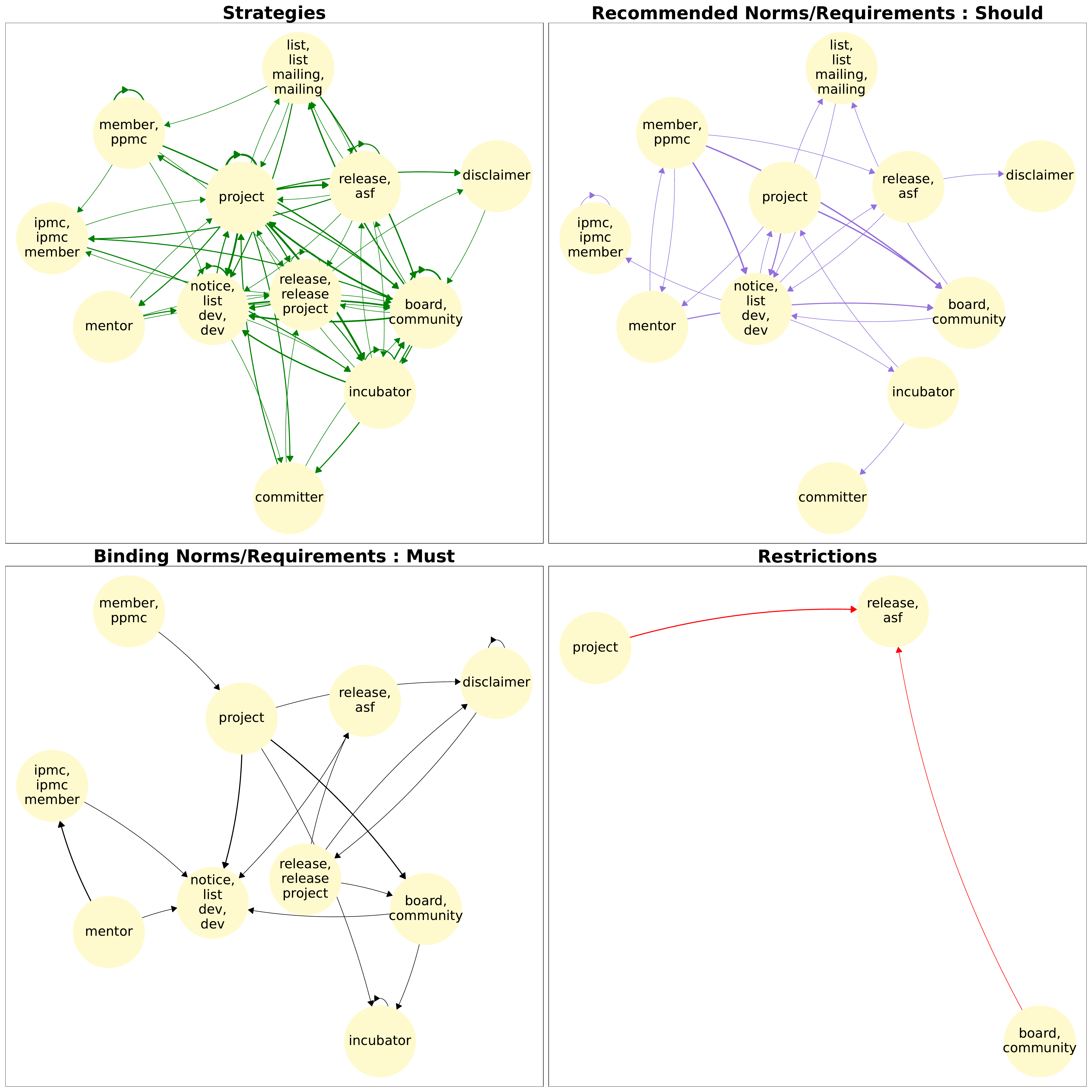}
\caption{Projects in the Apache Software Foundation Incubator often comprise volunteer developers and are mostly directed by strategies, with fewer strong regulations or restrictions. The network was generated by parsing policy constituents and aggregating similar actors/objects into nodes. Edges are directed from actors to objects and are logarithmically weighted by the number of policies between the pair. Node labels are the top representative words from each cluster. Projects are nodal actors and objects and are subject to certain recommended and binding practices towards their communities, software releases, the Incubator, Foundation board, Management Committees (PPMC/IPMC), and mentors. Notably, there are very few restrictions, and they are only applicable with regard to project releases.}

\label{fig:network}
\vspace*{-10pt}
\end{figure}

%  \begin{table}[!ht]
%  \centering
%  \begin{tabularx}{\columnwidth}{
%   |p{\dimexpr.4\columnwidth-2\tabcolsep-1.3333\arrayrulewidth}% column 1
%   |p{\dimexpr.15\columnwidth-2\tabcolsep-1.3333\arrayrulewidth}% column 2
%   |p{\dimexpr.15\columnwidth-2\tabcolsep-1.3333\arrayrulewidth}% column 2
%   |p{\dimexpr.15\columnwidth-2\tabcolsep-1.3333\arrayrulewidth}% column 2
%   |p{\dimexpr.15\columnwidth-2\tabcolsep-1.3333\arrayrulewidth}|% column 2
%   }
%         \hline
%         Dataset & Attribute (F1) & Object (F1) & Aim (F1) & Deontic (F1) \\ \hline
%         Food Policy Data (N = 398) & 0.71 &  0.57 &  0.82 &  0.94   \\ 
%         \hline
%         % Food Policy Data (Rice et al.) & 0.62 & \textbf{0.75} &   \textbf{0.85} & \textbf{0.97} \\ 
%         % \hline
%         Aquaculture Policy (N = 153) & 0.76 &  0.57 &  0.81 & 0.93  \\ 
%         \hline
%          National Organic Policy (N = 835) & 0.86 & 0.49 & 0.84 & 0.94  \\ 
%         \hline
%         % Community & Rule & Community & Rule & Similarity Score \\ 
%         % \hline
%         %  \hline
%     \end{tabularx}
%     \caption{Performance of Semantic Role Labeling for parsing Institutional Grammar Constituents}
%  \label{tab:val}
%  \vspace*{-10pt}
% \end{table}

 \subsubsection{Modeling and Visualizing Institutions}
 
The Institutional Analysis pipeline comprises three major NLP tasks. The first task involves coreference resolution (\textit{ABDICO\_coreferences}) using the implementation developed by Lee et al. ~\cite{lee-etal-2018-higher}. Coreferences are pairs of words referring to the same entity or object, e.g., '\textit{Paula} got a new sweater. \textit{She} loves it'. This module reads sections from policy documents and substitutes pronouns, articles and other determiners across sentences with the actual named entity they are referring. This preprocessing preserves valuable information and continuity of context between sentences and policy components when documents are tokenized into institutional statements and parsed. Policy constituents can be then parsed using \textit{ABDICO\_parsing}, which has been described in Section ~\ref{ig_parsing}.

The final module groups together the extracted components through their semantic similarity (\textit{ABDICO\_clustering}). This is particularly useful to discover patterns in the frequency and directionality of regulation between related institutional entities and instruments. It uses BERTopic~\cite{grootendorst2022bertopic}, a versatile topic modeling library that also uses sentence embeddings \cite{Reimers_Gurevych_2019} for text clustering, and can be extended to thematic categorization of policies or their constituents. IG components are used to derive qualitative inferences about institutions and even further inform downstream statistical and network analysis. Fig. ~\ref{fig:network} visualizes institutional interactions among OSS communities from the Apache Software Foundation Incubator \cite{Sen_Atkisson_Schweik_2022}. We only consider institutional statements where both objects and attributes were explicitly specified, and for the purposes of interpretable aggregation set the minimum topic size at 10 components. We adopt the SNR (Strategy-norm-requirements) taxonomy from Frey et al. ~\cite{frey_governing_2022}. Strategies are day-to-day operations or optional processes (can/may) that community members adopt to their needs, while norms and requirements are denoted by stronger deontics, such as should (recommended practices) and must (binding). Norms and requirements are similar, except the latter often explicitly lay out penalties and consequences for non-compliance. For the purposes of the current illustration, we consider policies broadly under this deontic-based scheme. Finally, we separately consider restrictions (cannot/may not, etc.), which are policies that forbid/discourage certain practices and are characterized by negation of the aim.  

\section{Further Work}
NLP4Gov is under active development as we continue to expand its capabilities along emerging directions in computational policy analysis, online collective action research, and language modeling. We are actively refining the existing features through domain adaptation of the pre-trained models with data from online communities and curation of validation benchmarks. Developments along policy decomposition could help realize granularity over a larger subset of the IG 2.0 \cite{frantz_institutional_2021} framework. Other pertinent components we intend to explore in the future are the "context" and “or else.” The context (C) spans the conditions and constraints upon which the policy is carried out, while Or-else (O) administers 'monitoring' and lays out consequences or penalties in the event the policy non-compliance.

The current implementation of our Institutional Grammar parser is designed to work for simple institutional statements, while interpretation and representation of policy concepts and their inter-dependencies is a cognitively involved process. Lately, large language models, through their emergent capabilities, intrinsic knowledge, and superior semantic understanding \cite{bubeck2023sparks}, hold immense promise for our intended goals. We intend to devise and demonstrate pedagogical prompting \cite{wei2022chain} methods that can leverage large language models. These can further help comprehend the nested structures and elucidate the underlying relationships between institutional instruments and entities. Low resource learning approaches, including LLM-based, \cite{brown2020language} can utilize limited expert annotations for confident and accurate predictions. Such methodologies, when adapted and demonstrated appropriately, are expected to be of particular interest to scholars who wish to operationalize different behavioral constructs over big data.

\section{Conclusion}
In the design of sociotechnical systems, it is often the case the advances in analyzing the specifying the "socio-" side lag behind "-technical" developments. In contrast to code, plain language articulations of formal governance structure suffer from all the ambiguities of human language, making written policy much more difficult to analyze and even design than the "code" part of sociotechnical law. Even where rigorous frameworks are available for policy analysis, they rely overwhelmingly on tedious hand-coding of subtle technical concepts, endangering inter-rater reliability, replicability, and generality, as quantitative studies of policies confine themselves to the analysis of single cases. 

With NLP4Gov, we provide computational, quantitative representations of written policies, increasing the rigor, scale, replicability, and accessibility of policy analysis advances.  As we have designed the toolkit, researchers with any level of programming proficiency can automatically perform a half dozen difficult policy analysis tasks, enabling semantic-level analysis and comparison of rules within and across institutions. 

By increasing access to tools for formally representing policy systems, we hope to empower information scientists to offer users more powerful policy design tools while empowering policy analysts and governance scholars with more powerful insights into what policies work and why. The need for such tools is especially pressing as technological advances reveal new threats and opportunities in the online social systems that structure our lives.

\section{Acknowledgement}
We would like to thank Saba Siddiki, Associate Professor of Public Administration and International Affairs, Syracuse University, Anamika Sen, Economics Ph.D. candidate at the University of Amherst, Massachusetts; and Likang Yin, Computer Science Ph.D. candidate at the University of California, Davis for data access. This work was supported by National Science Foundation grants \#2020751 and \#1917908.

\bibliographystyle{plain} 
\bibliography{main.bib}
\appendix
\section{Validation Data: Institutional Grammar}
\label{validation}

The Food Policy data \cite{siddiki2014assessing, rice_machine_2021} comprises 19 documents coded according to the Institutional Grammar (IG). We reconstructed polices from the word-level annotations used by Rice et al. (excluding punctuations). We also evaluated our approach on two other datasets. The Colorado aquaculture rules \cite{siddiki2012using} encompass 153 institutional statements. These, too, were developed by applying IG (ABDICO syntax \cite{frantz_institutional_2021}) to three regulatory documents governing aquaculture practices in the State of Colorado. These include the Colorado Aquaculture Act Statute (59 statements), the Rules Pertaining to the Administration and Enforcement of the Colorado Aquaculture Act (54 statements), and the section on fish health in the Colorado Division of Wildlife regulations (40 statements). Lastly, the National Organic Program regulations dataset \cite{carter2016integrating, siddiki2023understanding} comprises approximately 1078 IG-coded institutional statements. 

Despite the increasing popularity of systematic corpus-based policy research, the availability of annotated datasets for ML-based development/validation is still limited. These datasets were collected for diverse case studies, and each was annotated specific to particular research questions and analytical approaches. Moreover, their usage for automated evaluation of language technologies was particularly challenging due to subtle differences in styles and subjective discretion of the annotators. We hereby describe some additional pre-processing steps that were necessary to ensure uniform evaluation. 

For the National Organic Program and Aquaculture datasets, several constituents that were not explicitly specified in the statement itself were found to have been filled in by annotators through external sources of information. These entries were unsuitable for validating algorithms meant to extract information only from available text inputs. We performed a preliminary cleaning from all three datasets to exclude policies that were not coded (e.g., bulleted items) or only carried abstractive/implicit annotations for all ABDI constituents. We report the number of statements retained for evaluation in Table. ~\ref{tab:val}. For statements carrying one or more implicit ADBI labels, these specific constituents were further excluded at the time of evaluation. We also noticed and resolved some coding consistencies across the datasets prior to evaluation. We reduced verbs to their lemma root for matching (i.e., "Driving" was represented as "Drive") to account for differences in tense for annotations of 'Aim'. We further exclude non-informative words (stopwords) from word-matching, i.e., "The Committee" and "Committee" are treated as one and the same. 

\end{document}